\documentclass[12pt]{iopart}
\usepackage{graphicx}

\begin{document}

\title[Universality of miscible-immiscible phase separation dynamics in]{Universality of miscible-immiscible phase separation dynamics in two-component Bose-Einstein condensates}

\author{Xunda Jiang$^{1,2}$, Shuyuan Wu$^{1,2}$, Qinzhou Ye$^{1,2,3}$ and Chaohong Lee$^{1,2}$}

\address{$^1$ Laboratory of Quantum Engineering and Quantum Metrology, School of Physics and Astronomy, Sun Yat-Sen University (Zhuhai Campus), Zhuhai 519082, China}
\address{$^2$ Key Laboratory of Optoelectronic Materials and Technologies, Sun Yat-Sen University (Guangzhou Campus), Guangzhou 510275, China}
\address{$^3$ Guangdong Provincial Key Laboratory of Quantum Engineering and Quantum Materials,  School of Physics and Telecommunication Engineering, South China Normal University, Guangzhou 510006, China}
\ead{lichaoh2@mail.sysu.edu.cn}
\vspace{10pt}

\begin{abstract}
 We investigate the non-equilibrium dynamics across the miscible-immiscible phase separation in a binary mixture of Bose-Einstein condensates.
  The excitation spectra reveal that the Landau critical velocity vanishes at the critical point, where the superfluidity spontaneously breaks down.
  We analytically extract the dynamical critical exponent $z=2$ from the Landau critical velocity.
  Moreover, by simulating the real-time dynamics across the critical point, we find the average domain number and the average bifurcation delay show universal scaling laws with respect to the quench time.
  We then numerically extract the static correlation length critical exponent $v=1/2$ and the dynamical critical exponent $z=2$ according to Kibble-Zurek mechanism.
  The scaling exponents $(v=1/2, z=2)$ in the phase separation driven by quenching the atom-atom interaction are different from the ones $(v=1/2, z=1)$ in the phase separation driven by quenching the Rabi coupling strength [PRL \textbf{102}, 070401 (2009); PRL \textbf{107}, 230402 (2011)].
  Our study explores the connections between the spontaneous superfluidity breakdown and the spontaneous defect formation in the phase separation dynamics.
\end{abstract}

%
\vspace{2pc}
\noindent{\it Keywords}: universal dynamics, quantum phase separation, Bose-Einstein condensate, critical exponent
%
%
%
%

\section{Introduction}

The non-equilibrium dynamics of quantum phase transitions have attracted great interest in many branches of physics, including  cosmology, particle physics and condensed matter physics~\cite{Sachdev2011,Morikawa1995,Kibble1980}.
When a system is driven across a phase transition and enters a symmetry broken phase, one of the most nontrivial results is the  creation of topological defects, such as domains~\cite{Kibble1976,Lee2009,Davis2011,Davis2012,Xu2016,Swislocki2013, Hofmann2014,Wu2017,Navon2015,Ye2018}, vortices~\cite{Anderson2008,Su2013,Wu2016} and solitons~\cite{Zurek2010,Witkowska2011,Zurek2009}.
The possibility to engineer a quantum phase transition and recover its universality from topological defects is of great significance in non-equilibrium physics~\cite{Polkovnikov2011}.
In recent years, due to their high controllability and robust quantum coherence, atomic Bose-Einstein condensates (BECs) become an excellent candidate for exploring non-equilibrium dynamics across phase transitions~\cite{Sadler2006, Meldgin2016, Baumann2011, Bloch2008, Dziarmaga2010, Navon2015, Polkovnikov2011, Lamporesi2013, Anquez2016, Barnett2011, Klinder2015, Nicklas2015, Feng2018, Pelissetto2017, Coulamy2017, Clark2016,Mistakidis2018,Utesov2018,Francuz2016}.

In recent years, multi-component BECs have been widely investigated in both experiments~\cite{McCarron2011, Modugno2002, Wieman2008, Lin2011, Cornell1998, Tojo2010} and theories~\cite{Lee2009, Davis2011, Davis2012, Swislocki2013, Hofmann2014, Wu2016, Wu2017, Xu2016, Timmermans1998, Ao1998, Trippenbach2000, Takeuchi2013, Alexandrov2002, Esry1998, Saito2007}.
Up to now, great efforts have been made to create multi-component BECs with different atomic species~\cite{McCarron2011, Modugno2002}, isotopes~\cite{Wieman2008} or spin states~\cite{Lin2011,Cornell1998,Tojo2010}.
Multi-component BECs exhibit rich physics not accessible in a single-component BEC, including phase separation with symmetry breaking~\cite{Timmermans1998, Ao1998, Trippenbach2000, Esry1998, Lee2009, Alexandrov2002, Takeuchi2013}, Josephon oscillation~\cite{Williams1999} and domain walls~\cite{Davis2011, Davis2012, Swislocki2013, Hofmann2014, Wu2017, Xu2016}.
Remarkably, the phase separation in multi-component BECs has been observed in several experiments~\cite{Wieman2008, McCarron2011, Lin2011, Modugno2002, Tojo2010}.
Through controlling the intra- and inter-component interaction via Feshbach resonance~\cite{Courteille1998, Cornish2000, Inouye1998}, multi-component BECs offer an ideal test bed to study the non-equilibrium physics of phase separation.
However, there is few work on non-equilibrium dynamics in multi-component BECs, in particular, the dynamics of phase transition is still unclear.

In this paper, we investigate the non-equilibrium  dynamics of phase separation in a binary mixture of atomic BECs.
When the system is driven across the critical point at a finite rate, the critical dynamics across a miscible-immiscible phase transition is studied.
Through calculating the Bogoliubov excitation spectrum, we find that the Landau critical velocity vanishes and the superfluidity breaks down at the critical point,
and we analytically extract the dynamical critical exponent  from the Landau critical velocity.
To show how the non-equilibrium dynamics appears, we numerically simulate the real-time dynamics in quench process, in which the intra-component interaction strength is linearly swept through the critical point.
From the non-equilibrium dynamics far from the critical point, we numerically extract two universal scalings for the average domain number and the average bifurcation delay with respect to the quench rate, and we find the critical exponents derived from the numerical results consist well with the analytical exponents.
These scaling exponents in the phase separation induced by tuning the atom-atom interaction are different from the ones ~\cite{Lee2009,Davis2011,Davis2012,Xu2016}, in which the phase separation is induced by tuning the Rabi coupling strength.

The paper is organized as follows.
In Sec.~\uppercase\expandafter{\romannumeral2}, we describe the model and discuss its ground states.
In Sec.~\uppercase\expandafter{\romannumeral3}, we implement the  Bogoliubov-de Gennes (BdG) analysis and obtain the Landau critical velocity.
In Sec.~\uppercase\expandafter{\romannumeral4}, we analytically extract the Kibble-Zurek (KZ) scalings  and numerically simulate the real-time dynamics of the phase transition and  extract the universal scalings. Finally, we give a brief summary and discussion in Sec.~\uppercase\expandafter{\romannumeral5}.

\section{Model}

We consider a mixture of two weakly interacting BECs.
The system is described by the Hamiltonian $\hat{H}=\hat{H}_{0}+\hat{H}_{I}$ with the single-body part
\begin{equation}
\hat{H}_{0}=\int{dx}\sum_{j=1,2}\hat{\psi}^{\dagger}_{j}\left(x\right)
\left[-\frac{\hbar^{2}}{2m_{j}}\frac{\partial^{2}}{\partial x^{2}}+V\left(x\right)    \right]\hat{\psi}_{j}\left(x\right),     \label{H0}
\end{equation}
and the two-body part
\begin{eqnarray}
\hat{H}_{I}=&\int{dx} \left\{  \sum_{j=1,2}\left[  \frac{g_{jj}}{2}\hat{\psi}^{\dagger}_{j}\left(x\right)\hat{\psi}^{\dagger}_{j}
\left(x\right)\hat{\psi}_{j}\left(x\right)\hat{\psi}_{j}\left(x\right)  \right]\right\} \nonumber\\
&+\int{dx}\left\{g_{12}\hat{\psi}^{\dagger}_{1}\left(x\right)\hat{\psi}^{\dagger}_{2}
\left(x\right)\hat{\psi}_{2}\left(x\right)\hat{\psi}_{1}\left(x\right)\right\}.                                      \label{H1}
\end{eqnarray}
Here, $g_{jj}=4\pi\hbar^2 a_{jj}/m_{j}>0$ and $g_{12}=2\pi\hbar^2 a_{12}(m_{1}+m_{2})/(m_{1}m_{2})>0$ characterize the intra- and inter-component interactions, with $m_{j}$ being the mass of a atom in component $j$, $a_{jj}$ and $a_{12}$ respectively denoting the intra-component and inter-component s-wave scattering lengths.
In experiments, $g_{jj}$ and $g_{12}$ can be adjusted by using Feshbach resonance~\cite{Courteille1998, Cornish2000, Inouye1998}.  For simplicity, we only consider the homogeneous case, namely the external trapping potential $V(x)=0$ .

\begin{figure}[!htp]
  \centering\includegraphics[width=0.6\columnwidth]{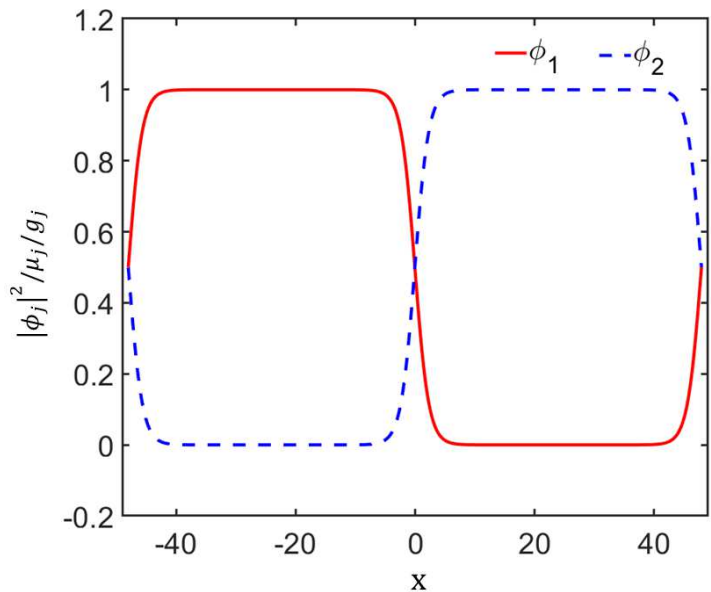}
  \caption{A typical ground state in the immiscible phase. The two BEC wave-functions occupy different spatial regions. The parameters are chosen as $N=2\times10^{6}$, $L=96$, $g_{11}=g_{12}=0.5$, and $g_{22}=0.49999$. }
  \label{Fig_Immiscible_Phase}
\end{figure}

In the mean-field (MF) theory, the system obeys the Gross-Pitaevskii equations (GPEs)
\begin{equation}
i\hbar \frac{\partial \psi_{j}}{\partial t}=\left[-\frac{\hbar^2}{2m_{j}} \frac{\partial^2}{\partial x^2}
+g_{jj}|\psi_{j}|^2+g_{12}|\psi_{3-j}|^2 \right]\psi_{j}. \label{GPE}
\end{equation}
The nature of the ground states is determined by the competition between the intra-component and inter-component interactions~\cite{Timmermans1998, Ao1998, Trippenbach2000}.
If the intra-component interaction dominates, i.e. $g_{11}g_{22}>g_{12}^2$, the energy is minimized when the two components occupy all available volume.
In such a miscible phase, the two BEC wave-functions coexist at all positions.
However, if the inter-component interaction dominates, i.e. $g_{11}g_{22}<g_{12}^2$, the energy of the system is minimized when the two BECs are separated in space.
In such an immiscible phase, the two BEC wave-functions occupy different spatial regions.
For the ground state in the immiscible phase, it is not easy to obtain the exact solution analytically.
However, by utilizing the imaginary time propagation method, we numerically solve Eq.~(\ref{GPE}) with a split-step method~\cite{Javanainen2006} with the Wick rotation $t=-i\tau$ and obtain the ground state under the period boundary condition.
In Fig.~\ref{Fig_Immiscible_Phase}, we show a typical ground state in the immiscible phase.
The two BEC wave-functions indeed occupy different space.
We observe an increase of the interface with the intra-component interaction strength $g_{jj}$.
If the intra-component interaction strength $ g_{jj} $ is stronger than a threshold $g_{jj}^{c}$ (or the inter-component interaction  $g_{12}$ is smaller than a threshold  $g_{12}^{c}$), the interface will cover the whole space, which is referred as miscible-immiscible phase transition.

To understand the critical dynamics near the phase transition, we introduce a dimensionless distance
\begin{equation}
\epsilon \left(t\right)=|g_{22}\left(t\right)-g_{22}^{c}|/g_{22}^{c},
\end{equation}
where $g_{22}^{c}=g_{12}^{2}/g_{11}$ is the critical point, and
\begin{equation}
 g_{22}\left(t\right)=g_{22}^{c}\left(1-t/\tau_{Q} \right)       \label{g22t}
\end{equation}
is a linearly quenched parameter with $\tau_{Q}$  being the quench time.
Below, we quench the intra-component interaction strength $g_{22}$ with different quench time $\tau_{Q}$ over several orders of magnitude.
Then, from our numerical simulation, we extract the relation between the number of topological defects, the bifurcation delay and the quench time $\tau_{Q}$.

\section{Bogoliubov excitation and spontaneous superfluidity breakdown near the critical point}

In this section, we analyse the Bogoliubov excitations near the critical point of phase separation and investigate the  spontaneous breakdown of superfluidity.
According to the Landau criterion, if the superfluid velocity is smaller than the Landau critical velocity, elementary excitations are prohibited due to the conservation of energy and momentum.
However, around the critical point, the Landau critical velocity  vanishes and so that elementary excitations appear spontaneously.

We implement a Bogoliubov analysis to obtain the excitation modes over the ground states.
In the miscible phase, the nonlinear Schr$\ddot{o}$dinger equations have an obvious homogenous solution $\rho_{j}=|\phi_{j}|^{2}=N_{j}/L$, with $\phi_{j}$  the ground state wave-function of the $j$-th component and $L$ the length of the system.
The chemical potentials are given as $\mu_{1}=g_{11}\rho_{1}+g_{12}\rho_{2}$ and $\mu_{2}=g_{22}\rho_{2}+g_{12}\rho_{1}$.
To derive the Bogoliubov excitation spectrum, we consider the perturbed ground state $($setting $\hbar=m=1)$
 \begin{equation}
 \psi_{j}\left(x,t \right)=[\phi_{j}+\delta \phi_{j}\left(x,t\right)]e^{-i\mu_{j}t}.   \label{Bogoliubov form1}
 \end{equation}
Inserting the perturbed ground state~(\ref{Bogoliubov form1}) into the GPE~(\ref{GPE}) and keeping the first-order terms, we obtain the linearized equations for the perturbations,
\begin{eqnarray}
i\frac{\partial \delta \phi_{1}}{\partial t}=&\left( -\frac{1}{2}\frac{\partial ^{2}}{\partial x^{2}}
+2g_{11}|\rho_{1}|+g_{12}|\rho_{2}|-\mu_{1} \right)\delta \phi_{1}\nonumber \\
&+g_{11}\phi_{1}^{2}\delta\phi_{1}^{*}+g_{12}\phi_{1}\phi_{2}^{*}\delta\phi_{2}
+g_{12}\phi_{1}\phi_{2}\delta\phi_{2}^{*},
\end{eqnarray}
\begin{eqnarray}
i\frac{\partial \delta \phi_{2}}{\partial t}=&\left( -\frac{1}{2}\frac{\partial ^{2}}{\partial x^{2}}
+2g_{22}|\rho_{2}|+g_{12}|\rho_{1}|-\mu_{2} \right)\delta \phi_{2}\nonumber \\
&+g_{22}\phi_{2}^{2}\delta\phi_{2}^{*}+g_{12}\phi_{2}\phi_{1}^{*}\delta\phi_{1}
+g_{12}\phi_{2}\phi_{1}\delta\phi_{1}^{*}.
 \label{Linearizing_equation 2}
\end{eqnarray}
The perturbations $\delta\phi_{1,2}$ can be written as
\begin{equation}
\left( {\begin{array}{*{20}{c}}
{\delta {\phi _1}}\\
{\delta {\phi _2}}
\end{array}} \right) = \left( {\begin{array}{*{20}{c}}
{{u_{1,q}}}\\
{{u_{2,q}}}
\end{array}} \right){e^{iqx - i\omega t}} + \left( {\begin{array}{*{20}{c}}
{{v_{1,q}^{*}}}\\
{{v_{2,q}^{*}}}
\end{array}} \right){e^{iqx + i\omega t}},  \label{fluctuation form1}
\end{equation}
in which $q$ is the excitation quasimomentum, $\omega$ is the excitation frequency, $u_{j,q}$ and $v_{j,q}$, $(j=1,2)$ are the Bogoliubov amplitudes.
Inserting Eq.~(\ref{fluctuation form1}) into the linearized equation Eq.~(\ref{Linearizing_equation 2}) and comparing the coefficients for the terms of $e^{iqx-i\omega t}$ and $e^{iqx+i\omega t}$, one obtains the BdG equations
\begin{equation}
\mathcal{M}(q) \mathbf{u}_{q}=\omega  \mathbf{u}_{q},
\end{equation}
\begin{equation}
\mathcal{M}(q) = \left( {\begin{array}{*{20}{c}}
{\hat{h}_{1}} &{g_{12}\phi_{1}\phi_{2}} &{g_{11}\phi_{1}^2} &{g_{12}\phi_{1}\phi_{2}}\\
{g_{12}\phi_{1}\phi_{2}} &{\hat{h}_{2}}&{g_{12}\phi_{1}\phi_{2}} &{g_{22}\phi_{2}^2}\\
{-g_{11}\phi_{1}^2} &{-g_{12}\phi_{1}\phi_{2}}&{-\hat{h}_{1}} &{-g_{12}\phi_{1}\phi_{2}}\\
{-g_{12}\phi_{1}\phi_{2}} &{-g_{22}\phi_{2}^2} &{-g_{12}\phi_{1}\phi_{2}} &{-\hat{h}_{2}}
\end{array}} \right),
\end{equation}
where $\mathbf{u}_{q} =\left(u_{1,q},u_{2,q},v_{1,q},v_{2,q}\right)^{T}$ and $\hat{h}_{j}=\frac{q^{2}}{2} +\sum_{k}g_{jk}|\phi_{k}|^2+g_{jj}|\phi_{j}|^2-\mu_{j}$.
Assuming that the two BECs have the same particle number,  $N_{1}=N_{2}=N/2$, we obtain the excitation spectrum by diagonalizing the matrix $\mathcal{M}(q) $ for each $q$
\begin{equation}
\omega_{\pm}^{2}=\epsilon_{0}\left(\epsilon_{0}+2\eta_{\pm} \right),    \label{excitation}
\end{equation}
with  $\epsilon_{0}=q^2/2$ and
\begin{equation}
\eta_{\pm}=\frac{\rho}{4}\left(g_{11}+g_{22}\pm \sqrt{\left( g_{11}-g_{22}  \right)^{2}+4g_{12}^{2}}   \right).
\end{equation}

At the critical point, the Landau critical velocity becomes
\begin{equation}
{v_L} = \mathop {\lim }\limits_{q \to 0} \left| {{\omega _ - }\left( q \right)/q} \right| = 0.
\end{equation}
The vanishing critical velocity at the critical point results in the spontaneous superfluidity breakdown and spontaneous elementary excitations.

\section{Kibble-Zurek scalings}

In this section, we extract two universal scalings from the Bogoliubov excitation in the miscible phase. As the softening of the phonon mode
near the critical point, we can extract  the dynamical critical exponent $z$ from the Landau critical velocity. Furthermore, we can make use of the Landau critical velocity to define correlation length, and extract another universal critical exponent $v$. At the same time,  we perform numerical simulations of non-equilibrium dynamics in the GPE~(\ref{GPE}).
Starting from the ground state in the miscible phase, according to Eq.~(\ref{g22t}), we linearly sweep the intra-component interaction strength $g_{22}$ to drive the system across the critical point $g_{22}^{c}$. When the evolution time $t$ increases from $t<0$ to $t>0$, the system goes from the miscible phase to the immiscible phase. Through analysing
the domains and bifurcation delay, we numerically obtain two universal critical exponents, which are consist well with the analytical exponents.
\begin{figure}[!htp]
  \centering\includegraphics[width=0.6\columnwidth]{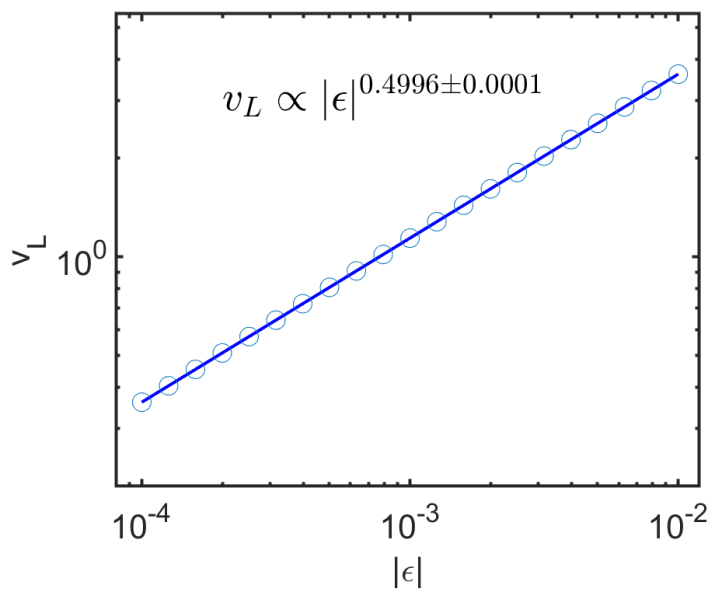}
  \caption{(Color online) The scaling of the Landau critical velocity $v_{L}$ versus $\left| \epsilon \right|$ near the critical point. The solid line
  is the linear fitting, which show that the $v_{L}$ has a power-law dependence on the   $\left| \epsilon \right|$. }
\label{Fig_landau_velocity}
\end{figure}

The dynamical critical exponent $z$ can be derived from the low-energy long-wave excitations.
At the critical point, $g_{22}=g_{22}^{c}=g_{12}^{2}/g_{11}$, the low-energy excitation in the long-wavelength limit behave as
\begin{equation}
\mathop {\lim }\limits_{q \to 0} {\omega _ - }\left( q \right) = {q^2}/2.
\end{equation}
Since ${\omega_-}\left( q \right) \propto {\left| q \right|^z}$ as $q \to 0$~\cite{Sachdev2011,Robinson2011,Polkovnikov2011}, we have the dynamical critical exponent $z=2$.

The static correlation length critical exponent $v$ can be derived from the divergence of the correlation length $\xi$ near the critical point.
The Landau critical velocity $v_{L}$ provide a definition of the correlation length $\xi$ according to $\xi  = \hbar /\left( {m{v_L}} \right)$~\cite{Giorgini2008}.
Therefore, the Landau critical velocity $v_{L}$ should
have a power-law scaling behavior around the critical point as
\begin{equation}
{v_L} \propto {\xi ^{ - 1}} \propto {\left| \epsilon \right|^v}
\end{equation}
In Fig.~(\ref{Fig_landau_velocity}), we plot the $v_{L}$ for different $\left| \epsilon \right|$ near the critical point in a log-log coordinate. It shows clearly that $v_{L}$
has a power-law dependence on $\left| \epsilon \right|$, which can be expressed by $v_{L} \propto {\left| \epsilon \right|}^b$. Through linear fitting, we find
$b=0.4996\pm0.0001$ for the miscible phase. This indicates that the static correlation length critical exponent $v=1/2$.

In our simulations, we choose quench times $\tau_{Q}$ over four orders of magnitude and  perform 100 runs of simulation for each $\tau_{Q}$.
Since the quantum fluctuations that trigger the phase transition are ignored in the MF approximation~\cite{Saito2007}, we introduce random noises to the initial state and so that the dynamics of spontaneous topological defects can be studied by the MF  theory.
\begin{figure}[htp]
  \centering\includegraphics[width=0.7\columnwidth]{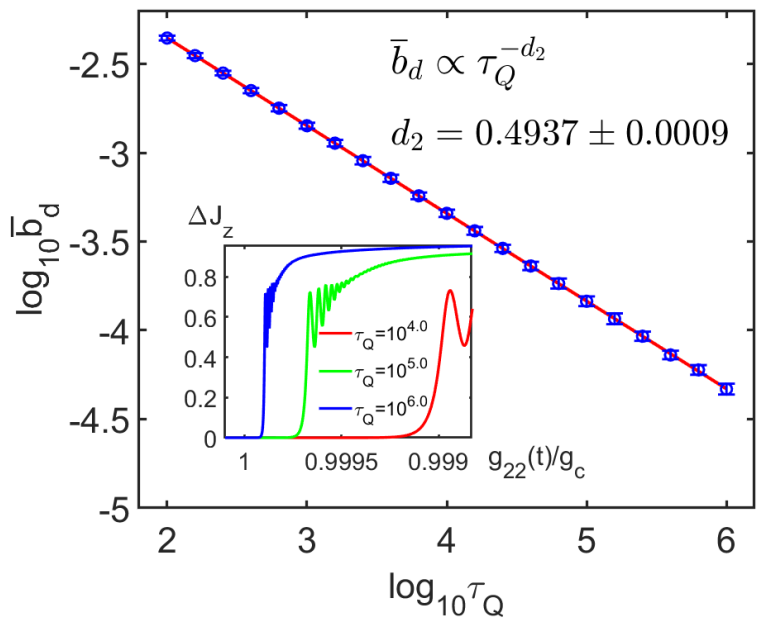}
  \caption{(Color online) Universal scaling of the average bifurcation delay $\overline{b}_{d}$ with respect to the quench time $\tau_{Q}$.
  The inset shows the growth of spin fluctuation $\Delta J_{z}$ after the system crossing the critical point with three typical quench times $\tau_{Q}$.
  The bifurcation delay is given as $\overline{b}_{d}=|g_{22}^{*}-g_{c}|$ with $g_{22}^{*}$ where $\Delta J_{z}$ reaches a small nonzero value 0.05.
  The error bars correspond to the standard deviation of 100 runs. The parameters are chosen as $N=2\times10^{6}$, $L=96$, and $g_{11}=g_{12}=0.5$. }
\label{Fig_Bifurcation_delay}
\end{figure}
As the interaction strength $g_{22}$ is quenched according to Eq.~(\ref{g22t}), unlike the static bifurcation, in which the bifurcation exactly occurs at the critical point, the dynamical bifurcation takes place after the system crossing the critical point.

The bifurcation delay, ${b_d} = \left| {{g_{22}^{*}} - g_{22}^c} \right|\propto {\left|\hat{ \epsilon} \right|}$, is obtained by analyzing spatial fluctuation of the local spin polarization
\begin{equation}
\Delta {J_z} = \sqrt {\frac{1}{L}\int {J_z^2\left( x \right)dx}
- {{\left[ {\frac{1}{L}\int {{J_z}\left( x \right)dx} } \right]}^2}},
\end{equation}
with $J_{z}=[n_{1}(x)-n_{2}(x)]/[n_{1}(x)+n_{2}(x)]$.
Before the bifurcation, there is no spatial fluctuation, i.e. $\Delta {J_z} =0$.
The dynamical bifurcation occurs at $g_{22}^{*}$ where  $\Delta{J_z}$ reaches a small nonzero value $\delta_{J}$, which is chosen as 0.05 in our calculations.
Actually, based upon our calculations, similar conclusions can be obtained for other small $\delta_{J}$ between $0.05$ to $0.2$.

\begin{figure}[!htp]
  \centering\includegraphics[width=0.7\columnwidth]{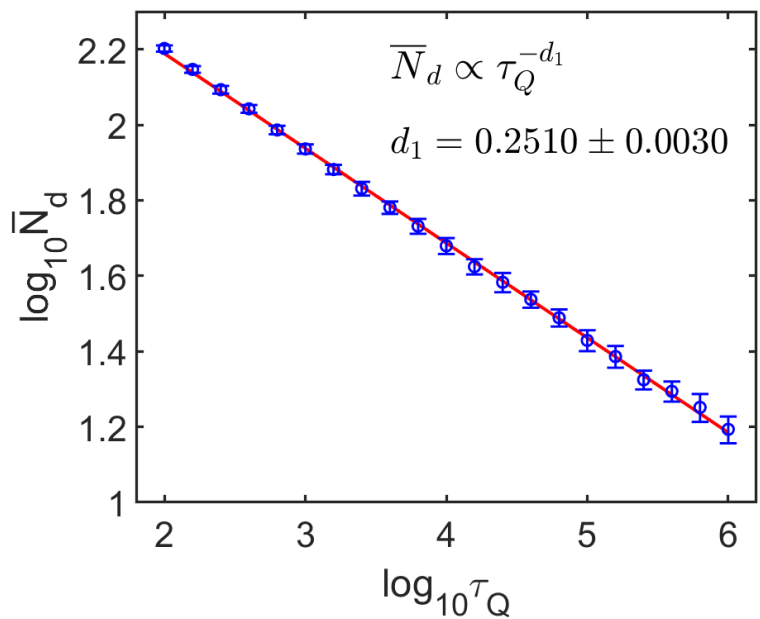}
  \caption{(Color online) Universal scaling of the average domain number $\overline{N}_{d}$ with respect to the quench time $\tau_{Q}$.
  The number of domains are counted at the end of evolution when the domains become stable.
  The examples of domains-formation dynamics are shown in Fig.~(\ref{Fig_Domian_formation}).
  The error bars correspond to standard deviation of 100 runs.
  The parameters are the same ones for Fig.~(\ref{Fig_Bifurcation_delay}).}
  \label{Fig_Domain_scaling}
\end{figure}

In Fig.~(\ref{Fig_Bifurcation_delay}), we show the dependence of the bifurcation delay on the quench time.
Clearly, there is a significant delay between the growth of local spin fluctuation and the static phase transition, see the inset of Fig.~(\ref{Fig_Bifurcation_delay}).
For a smaller quench time, the system has a larger bifurcation delay $b_{d}$. In our numerical results, it is clearly illustrate that
the average bifurcation delay $\overline{b_{d}}$  follows a power-law scaling with respect to the quench time $\tau_{Q}$, $\overline{b}_{d}\propto \tau_{Q}^{-d_{2}}$ with the scaling exponent $d_{2}=0.4937$. According to the KZ mechanism, we have
\begin{equation}
\hat{\epsilon} \propto \tau_{Q}^{-\frac{1}{1+vz}}=\tau_{Q}^{-0.4937}.   \label{bd_exponent_equation}
\end{equation}

\begin{figure*}[htp]
  \centering\includegraphics[width=1\columnwidth]{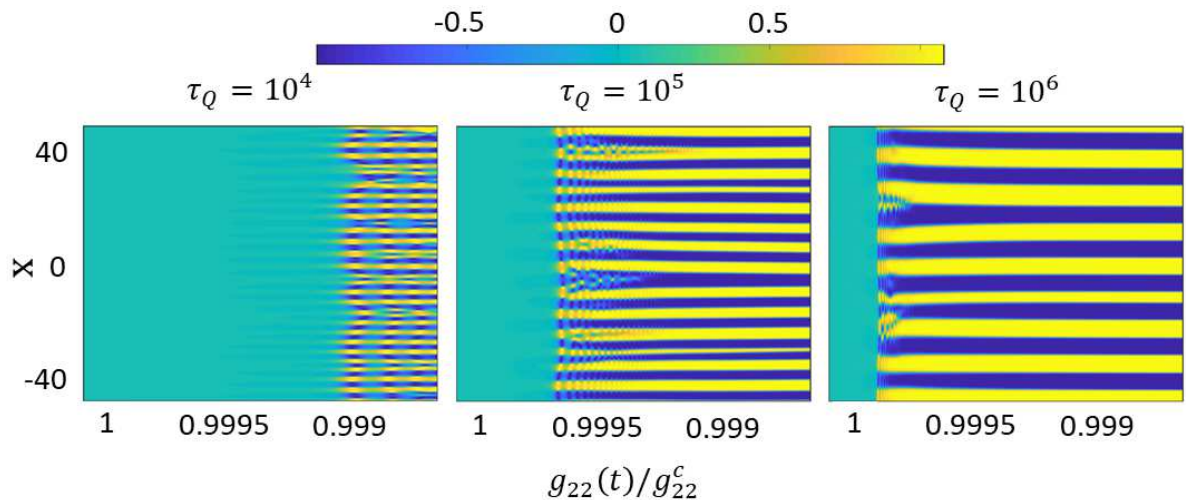}
  \caption{The domain formation in the real-time dynamics for different quench times $\tau_{Q}$. When the system is quenched from miscible to immiscible phases at a finite $\tau_{Q}$, due to the vanishing of Landau critical velocity at the critical point, elementary excitations emerge and domains are spontaneously created. The parameters are chosen as $N=2\times10^{6}$, $L=96$, $g_{11}=g_{12}=0.5$, and $\tau_{Q}=\{10^{4}, 10^{5}, 10^{6}\}$.}
  \label{Fig_Domian_formation}
\end{figure*}

In order to extract the critical exponents, we also analyse the universal scaling of domain number $N_{d}$ versus the quench time $\tau_{Q}$.
Typical examples of domain formation dynamics for different quench times are illustrated in Fig.~(\ref{Fig_Domian_formation}).
One can find that the average domain size increases with the quench time.
In each run, we count the number of domains $N_{d}$ by identifying the number of zero crossings of $J_{z}$ at the end of evolution, when the domain shape becomes stable.
The number of domains for different $\tau_{Q}$ are shown in Fig.~(\ref{Fig_Domain_scaling}).
We observe that the average domain number follows a power-law scaling $\overline{N_{d}}\propto \tau_{Q}^{-d_{1}}$ with the scaling exponent $d_{1}=0.2510$.
 According to the KZ mechanism, we have
 \begin{equation}
N_{d}\propto \tau_{Q}^{-\frac{v}{1+vz}}=\tau_{Q}^{-0.2510}   \label{domain_exponent_equation}
\end{equation}

Combining Eq.~(\ref{bd_exponent_equation}) and Eq.~(\ref{domain_exponent_equation}), we obtain the static correlation length critical exponent $v=0.5084$ and the dynamical critical exponent $z=2.0172$.
These exponents well consist with the ones $(v=1/2, z=2)$ obtained from the Landau critical velocity and the correlation length.
These critical exponents are in the same universal class as in the previous work~\cite{Ye2018,Wu2017,Wu2016,Zurek2010,Clark2016}. However, in a binary BEC with Rabi coupling~\cite{Lee2009,Davis2011,Davis2012,Xu2016} across the phase separation driven by quenching the Rabi coupling strength, the exponents are given as $(v=1/2,z=1)$.
This means that, although the phase separation can be induced by quenching either the Rabi coupling or the atom-atom interaction, they belong to different universal classes.



\section{Summary and discussion}

In summary, we have investigated the non-equilibrium dynamics across a phase separation transition in a mixture of two-component BECs.
Through analyzing the Bogoliubov excitation spectrum, we find that the Landau critical velocity vanishes at the critical point, which results in the spontaneous superfluidity breakdown and creation of elementary excitations, and we extract the critical exponents from the Landau critical velocity and the correlation length near the phase transition.
When the system is driven across the critical point with a finite quench rate, the domains are spontaneously created due to the vanishing of the spontaneous superfluidity breakdown.
On the other hand, by numerically simulating the GPEs, we find that domains appear after the system crossing the critical point  $g_{22}^{c}$, and we count the domains number at the end of evolution.
Through quenching the system with various $\tau_{Q}$, we find two universal scalings of the average bifurcation delay and average domain number versus the quench rate.
We then numerically extract two critical exponents $(v=1/2,z=2)$ from two universal scalings.

Although all this work and the previous ones~\cite{Lee2009,Davis2011,Davis2012,Xu2016} study miscible-immiscible transitions, there are significant differences between their KZ scalings.
Firstly, the miscible-immiscible transitions are induced by different sources. In the previous works~\cite{Lee2009,Davis2011,Davis2012,Xu2016}, the miscible-immiscible transitions are controlled by the Rabi coupling. On the other hand, in our system, the miscible-immiscible transition is controlled by the interaction coefficient.
Secondly, the pictures for understanding the KZ mechanism are different. In the previous works~\cite{Lee2009,Davis2011,Davis2012,Xu2016}, the KZ mechanism is understood via the gapless excitations. However, in our manuscript, the KZ mechanism is understood via the divergence of correlation length derived from the Landau critical velocity. So far, we analytically extract the dynamical critical exponent from the Landau critical velocity and determine the static correlation length critical exponent from the correlation length derived from the Landau critical velocity.
Thirdly, and most importantly, the KZ scaling exponents are different. In our manuscript, the scaling exponents are given as (v = 1/2, z = 2). However, in the previous works~\cite{Lee2009,Davis2011,Davis2012,Xu2016}, the scaling exponents are given as (v = 1/2, z = 1).
This means that the miscible-immiscible transition driven by quenching the atom-atom interaction and the miscible-immiscible transition driven by quenching the Rabi coupling belong to different universal classes.

Based upon current available techniques for atomic BECs, it is possible to probe the above universal scalings.
The mixture of two-component BECs can be prepared with different atomic species~\cite{McCarron2011, Modugno2002}, isotopes~\cite{Wieman2008} or spin states~\cite{Lin2011,Cornell1998,Tojo2010}.
The phase separation has been observed in several experiments~\cite{Wieman2008,Tojo2010} and the high tunability of the interaction strength makes the quenching across the phase separation transition possible.
Furthermore, the bifurcation delay and the size of domains can be extracted by measuring the local spin polarization via the time-of-flight.

\begin{ack}

This work was supported by  the National Natural Science Foundation of China (Grants No. 11374375, No. 11574405).

\end{ack}

\section*{References}
%
%
%
\bibliographystyle{unsrt}
\bibliography{S_KZM_Reference2}

%

\end{document}